\documentclass[12pt]{article}
\usepackage{comment}
\usepackage{multirow}
\usepackage[T1]{fontenc}    
\usepackage[utf8]{inputenc} 
\usepackage[dvipsnames]{xcolor} 
\usepackage{amsmath}
\usepackage[titletoc]{appendix}
\usepackage{array}
\usepackage[english]{babel}
\usepackage{bm}
\usepackage{booktabs}
\usepackage{caption} 
\usepackage{enumitem}
\usepackage{float}
\usepackage{gensymb}
\usepackage[left=2.5cm,right=2.5cm,top=2.5cm,bottom=2.5cm]{geometry}
\usepackage{graphicx}
\usepackage[colorlinks,bookmarks=false,linkcolor=blue,urlcolor=blue,citecolor=blue]{hyperref}
\usepackage{lipsum}
\usepackage{parskip}
\usepackage{setspace} 
\usepackage{sidecap}
\usepackage{siunitx}
\usepackage{subcaption}
\usepackage{titlesec}
\usepackage[nottoc]{tocbibind}
\usepackage[normalem]{ulem}
\usepackage{wasysym}
\usepackage{wrapfig}
\usepackage[semicolon]{natbib}
\usepackage[export]{adjustbox}
\usepackage{physics}
\usepackage[utf8]{inputenc}
\usepackage{tikz}

\def \be {\begin{equation}}
\def \ee {\end{equation}}

\def \8 {\infty}

\usepackage{amsmath}
\usepackage{bm}
\usepackage{geometry}
\title{Seismic noise interferometry for phase transmission fibre optics}
\author{Sixtine Dromigny $^1$, Daniel Bowden $^2$, \\
Sebastian Noe $^2$, Dominik Husmann $^3$, Andreas Fichtner $^2$\\
$^1$ Department of Earth Sciences, University of Oxford, United Kingdom\\
$^2$ Department of Earth and Planetary Sciences, ETH Zurich, Switzerland\\
$^3$ Federal Institute of Metrology, METAS, Switzerland}
\date{}

\begin{document}
\maketitle

\section*{Abstract}

Similar to Distributed Acoustic Sensing (DAS), phase transmission fibre optics allows for large bandwidth seismic data measurements using fibre-optic cables. However, while the application range of DAS is limited to tens of kilometres, phase transmission fibre optics has an application range that can go up to thousands of kilometres. This new method has been shown as an effective method to record earthquakes, but its ability to record ambient seismic noise that can be used for seismic imaging and tomography is still up for question, and will be analysed in this work. We provide the theoretical foundation for the interpretation of seismic noise autocorrelations and interferometry from phase transmission fibre optics. Further, we test the model on actual phase transmission data sourced from a phase-stabilised optical frequency network in Switzerland. There, the phase stabilisation scheme measures and compensates noise on the optical phase caused by distortions of the fibre. We analyse the autocorrelation of the measured phase noise correction and explore potential interpretations by comparing it with the autocorrelation of a synthetically computed phase noise correction. This comparison is challenging due to two factors: the intricate cable geometry increases the computational cost of generating synthetic data, and the precise location and geometry of the cable are uncertain. Despite these difficulties, we believe that when applied to a different dataset, this approach could enable seismic tomography with ambient noise interferometry using a long-range fibre-optic sensing device.

\section*{Keywords}
Theoretical seismology, Wave propagation, Seismic tomography, Seismic instruments, Computational seismology, Ambient seismic noise interferometry, Fibre-optic sensing

\section{Introduction}

Seismic ambient noise (SAN) is a low-intensity signal recorded by seismometers. It refers to a quasi random wavefield excited by wind, human activities or interaction of ocean waves with the ocean bottom \citep[e.g.,][]{Ardhuin_2011,Gualtieri_2013,Gimbert_2016,Gualtieri_2019,nakata2019seismic}.  Seismic noise can also be used to infer inner Earth structures through interferometry \citep[e.g.,][]{Shapiro_2005,saygin2017retrieval,Stehly_2009,Boue_2013,Lin_2008,Mordret_2013,Sager_2018b,Sager_2020}, which is especially useful in regions with low earthquake occurrences. Cross-correlating signals from two recording stations enables seismic tomography, as the cross-correlation between the signals at these stations approximates the subsurface response to an impulsive source. Mathematically, this is represented as the Green's function, where one station acts as the source and the other as the receiver \citep[e.g.,][]{Lobkis_2001,Weaver_2004,Wapenaar_2004,Wapenaar_2006,curtis2006seismic,snieder2010imaging,Fichtner_2019b}. This approximation holds under the assumption that either the noise sources are homogeneously distributed and uncorrelated, or that the different wave propagation modes are uncorrelated and and equipartitioned.

As an emerging seismic measuring technology, Distributed Acoustic Sensing (DAS) also records ambient seismic noise. This technique relies on the scattering of light waves within an optical fibre, resulting from the light wave encountering heterogeneities as it propagates through the fibre. When the fibre is strained, the signature of back-scattered light changes, and one can relate the phase delay to the strain caused on the cable by the ground motion with a high spatial resolution along the fibre \citep[e.g.,][]{daley2013field,Hartog_2017,Lindsey_2021}. Thanks to its broad frequency response \citep{lindsey2020broadband,Paitz_2021}, DAS has found widespread application in urban seismology \citep[e.g.,][]{Spica_2020,Biondi_2017,Dou_2017,Smolinski_2024}, volcanology \citep[e.g.,][]{Currenti_2022,Jousset_2022,Klaasen_2021,Klaasen_2023}, glaciology \citep[e.g.,][]{Brisbourne_2021,Booth_2020,Fichtner_2023,Fichtner_2023b} and many other domains.

Unfortunately, the attenuation of the back-scattered signal limits the interrogation distance DAS. Current DAS interrogators can mostly not record useful data beyond 100 km. In contrast to DAS, new techniques are being developed based on the deformation-induced phase changes of the forward transmitted light signal \citep[e.g.,][]{marra2018ultrastable,Bogris_2022,noe2023long,donadello2024seismic}. Strain-induced elongation or contraction of the fibre alter its length and refractive index, which in turn affects the propagation time of light waves. Consequently, ground motion can be measured similarly to DAS by relating the seismic-induced strain on the fibre to variations in light wave travel time. Moreover, these phase-transmission based methods enable a greater propagation range, as the forward-propagated light can traverse the cable, and the intensity of the forward-propagating light is inherently higher than that of the back-scattered light. Additionally, amplifiers can be employed to offset any loss in intensity. Thus, measurements could be collected over a much longer distance, for example to measure ground motions on the sea floor over thousands of kilometres of pre-existing telecommunication cables \citep{marra2018ultrastable}.

The work of \cite{fichtner2022theory} demonstrates how phase variations in signals transmitted through long fibres are related to the seismic wavefield. However, further theoretical advancements are necessary to adapt current seismic methods for recordings that represent strain distributed over extended distances. Moreover, the fibre provides only a single measuring channel, introducing an additional challenge.
One important aspect of the measurement pointed out by \cite{fichtner2022theory} is that the measurement sensitivity depends on the curvature of the fibre. Along a perfectly straight segment of fibre the change in length and refractive index will average out and no signal would be recorded. To infer the phase changes along the whole fibre, we would need to integrate the phase changes along the fibre, which would lead to a loss of information \citep{fichtner2022sensitivity}.
\cite{fichtner2022theory} hence provide a formalism that allows to extract spatially distributed measurements from of the integrated ones. High-curvature segments along the fibre may serve as local sensors, thereby localising the measurement along the fibre. This relation between fibre curvature and spatial resolution has effectively been used for earthquake detection and location \citep{noe2023long,mueller2024earthquake,donadello2024seismic}, but it has not been yet applied to SAN recordings.

While for traditional SAN interferometry, one analyses the cross-correlation between different seismic stations, in the case of the phase transmission fibre measurement, only a single time series of integrated strain is available, and so instead we turn to the autocorrelation of that signal with itself. To evaluate the transmitted phase signal for conducting seismic tomography using seismic noise recordings, this paper will first analytically express the autocorrelation of the transmitted phase changes. We will demonstrate that this autocorrelation equation provides a deterministic expression of the signal, even though it is derived from noise sources of quasi random nature. We then provide an analytical expression of the autocorrelation equation for a simplified fibre geometry, in order to illustrate the potential of the method to achieve a spatial coverage that may enable seismic tomography.

We subsequently test our framework using actual data from a phase noise cancellation system \citep{Husmann:21}. By autocorrelating this noise cancellation signal, we expect to be able to perform seismic tomography. To test our hypothesis, we calculate the expected recorded strain of the cable for ambient seismic noise in a set seismic velocity model of Switzerland using the spectral-element solver Salvus \citep{afanasiev2019modular}, and compare the simulated data set to actual data provided by METAS. Unfortunately, the comparison is difficult to interpret on this scale, for several reasons discussed later in the paper. 

\section{Theoretical derivation of the autocorrelation equation}\label{sec:1}

\cite{fichtner2022theory} present an equation describing the phase transmission of the light signal through a single optical fibre,

\begin{equation}
    \dot \phi(t)= \int_{s=0}^{L} - \frac{\omega}{c_0} \pdv{}{s} \left[r_{eff}(s) \bm{e}(s)\right]^T \pdv{}{t} \bm{u} (s,t) ds\,,
    \label{eq:phase}
\end{equation}
with $s$ the distance along the fibre, $\dot \phi(t)$ the time derivative of the transmitted phase changes, 
$\omega$ the circular frequency of the light signal,  
$c_0$ the speed of light in vacuum, 
$r_\text{eff}(s)$ the effective refractive index along the fibre, 
$\textbf{e}(s)$ the tangent unit vector along the fibre, and $ \bm{u} (s,t)$ the displacement field. In \cite{fichtner2022theory}, equation (\ref{eq:phase}) is simplified as follows,
\begin{equation}
    \dot \phi(t) = \int_{s=0}^{L} \bm{a}(s)^T \pdv{}{t} \bm{u} (s,t)d\bm{s} \,,
    \label{eq:a_add}
\end{equation}
where $\bm{a}(s)=\frac{\omega}{c_0} \pdv{}{s} \left[r_{eff}(s) \bm{e}(s)\right] $. We define the autocorrelation of the phase transmission as
\begin{equation}
    c(\tau) = \int_{-\infty}^{\infty} \dot \phi(t)  \dot \phi(t + \tau) dt\,.
\end{equation}
In the Fourier domain, the autocorrelation can be written as a product,
\begin{equation}
    c(\omega) = -\omega^2  \phi^*(\omega)\phi(\omega)\,,
    \label{eq:correlation}
\end{equation}
where $*$ denotes complex conjugation. When we integrate the phase transmission time series from equation (\ref{eq:a_add}) and express it in the Fourier domain we obtain
\begin{equation}
    \phi(\omega) = \int_{s=0}^{L} \bm{a}(s) \bm{u} (s,\omega)d\bm{s}\,. 
    \label{eq:phaseFourier}
\end{equation}
We then apply the representation theorem \citep[e.g.,][]{aki2002quantitative} to express a displacement field $u_i (\bm{x},t)$ in terms of the Green's function $G_{in} (\bm{x}, \bm{x_0}, t-t_0)$ and the spatially extended force $f_n(\bm{x_0},t_0)$,
\begin{equation}
    u_i (\bm{x},t) = \int_{-\infty}^{\infty} \int_{-\infty}^{\infty} G_{in} (\bm{x}, \bm{x_0}, t-t_0) f_n(\bm{x_0},t_0) d\bm{x_0}dt_0\,.
    \label{eq:green}
\end{equation}
When we express equation (\ref{eq:green}) in the Fourier domain, the following equation results,
\begin{equation}
    u_i (\bm{x},\omega) = \int_{-\infty}^{\infty}  G_{in} (\bm{x}, \bm{x_0}, \omega) f_n(\bm{x_0},\omega) d\bm{x_0}\,.
    \label{eq:green_Four}
\end{equation}
We now substitute equation (\ref{eq:green_Four}) into equation (\ref{eq:phaseFourier}), which leads to
\begin{equation}
    \phi(\omega) = \int_{\bm{s}=0}^{L} a_i(s) \left[\int_{-\infty}^{\infty} G_{in} (\bm{x}(s), \bm{x_0}, \omega) f_n(\bm{x_0},\omega) d\bm{x_0}\right]ds\,.
    \label{eq:phaseFourier_rep}
\end{equation}
For simplicity, we substitute $\bm{x}(s)$ with $s$, which corresponds to the path along the fibre. We now express the autocorrelation of the previously found phase transmission expression (see equations (\ref{eq:correlation}) and (\ref{eq:phaseFourier_rep})),
\begin{multline}
    c(\omega)=-\omega^2  \left[\int_{-\infty}^{\infty}\int_{s=0}^{L} a_i(s) G_{in}^* (s, \bm{x_0}, \omega)  f_n^*(\bm{x_0},\omega) d\bm{x_0}ds \right]\\
    \left[\int_{-\infty}^{\infty}\int_{s'=0}^{L} a_j(s') G_{jm} (s', \bm{x_0}', \omega)  f_m(\bm{x_0}',\omega) d\bm{x_0}'ds'\right]\,.
    \label{eq:fast_endergebnis}
\end{multline}
We consider the following definition of the power-spectral density (PSD) of noise sources at two different source locations, $\bm{x_0}$ and $\bm{x_0}'$, denoted as $S_{nm}(\bm{x_0},\bm{x_0}')$,
\begin{equation}
    S_{nm}(\bm{x_0},\bm{x_0}') = \langle f_n^*(\bm{x_0},\omega)  f_m(\bm{x_0}',\omega)\rangle\,,
    \label{eq:PSD}
\end{equation}
where $\langle . \rangle$ denotes the ensemble average over the noise sources. As we assume that neighbouring noise sources are uncorrelated in time, we can rewrite equation (\ref{eq:PSD}) as,
\begin{equation}
    S_{nm}(\bm{x_0},\bm{x_0}') = S_{nm}(\bm{x_0})\delta(\bm{x_0}-\bm{x_0}')\,.
    \label{eq:PSD_uncorr}
\end{equation}
Substituting equation (\ref{eq:PSD}) into equation (\ref{eq:PSD_uncorr}), we obtain the following expression for the PSD of uncorrelated noise sources,
\begin{equation}
    S_{nm}(\bm{x_0})\delta(\bm{x_0}-\bm{x_0}') = \langle f_n^*(\bm{x_0},\omega)  f_m(\bm{x_0}',\omega) \rangle\,.
    \label{eq:PSD_def}
\end{equation}
We then replace $ \langle f_n^*(\bm{x_0},\omega)  f_m(\bm{x_0}',\omega) \rangle$ in equation (\ref{eq:fast_endergebnis}) with its definition from equation (\ref{eq:PSD_def}) and thus achieve our end result for the autocorrelation depending on the noise sources,
\begin{equation}
    c(\omega)=-\omega^2 \int_{s=0}^{L} \int_{s'=0}^{L} a_i(s)  a_j(s') \left[ \int G_{in}^* (s, \bm{x_0}, \omega)  G_{jm} (s', \bm{x_0}, \omega)  S_{nm}(\bm{x_0}) d\bm{x_0} \right]dsds'\,.
    \label{eq:endergebnis}
\end{equation}
Equation (\ref{eq:endergebnis}) involves a numerically challenging triple integration. To address this, we replace the cross-correlation of noise wavefields in equation (\ref{eq:endergebnis}) by the commonly utilised Green's function approximation \citep[e.g.,][]{Lobkis_2001,Weaver_2004,Wapenaar_2004,Wapenaar_2006,curtis2006seismic,snieder2010imaging,Fichtner_2019b},
\begin{equation}
    \int_{-\infty}^{\infty} G_{in}^* (s, \bm{x_0}, \omega)  G_{jm} (s', \bm{x_0}, \omega)  S_{nm}(\bm{x_0}) d\bm{x_0} \propto G_{ij}(s,s',\omega)\,.
    \label{eq:approx}
\end{equation}
Equation (\ref{eq:approx}) removes not only one of the integrals but also the dependence on the noise source PSD. However, equation (\ref{eq:approx}) is strictly valid only under the condition that the noise sources originating at different points in space $\bm{x_0}$ are distributed homogeneously throughout the domain. We substitute equation (\ref{eq:approx}) into equation (\ref{eq:endergebnis}) to obtain
\begin{equation}
    c(\omega) \propto -\omega^2 \int_{s=0}^{L} \int_{s'=0}^{L} a_i(s)  a_j(s') G_{ij}(s,s',\omega) dsds'\,.
    \label{eq:implement}
\end{equation}
In equation (\ref{eq:implement}), we can see that the autocorrelation of the phase transmission signal in the Fourier domain is proportional to the curvature of the fibre multiplied by the Green's functions originating at these points. In the following section, we will focus on the special case where the fibre has a discrete number of kinks, i.e., points of infinite curvature. This will serve to highlight the proportionality between the autocorrelation and and the curvature of the fibre more clearly.

\section{Conceptual example}

\begin{figure}
    \includegraphics[width=1\linewidth]{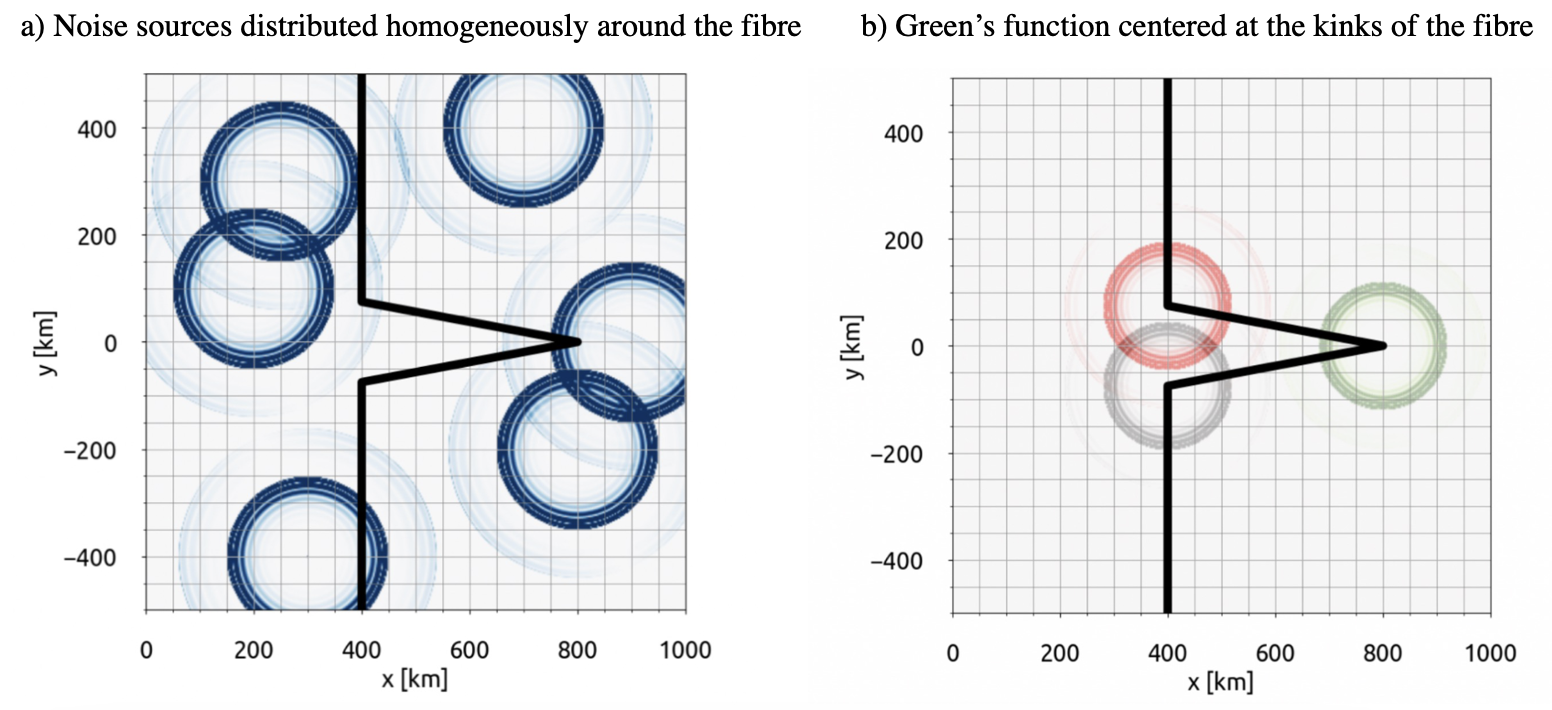}
    \caption[Illustrated example of a triangular fibre geometry]{Conceptual example of phase transmission ambient noise interferometry. The fibre is drawn in black in (a) and (b). (a) Different sources distributed around the fibre excite P and S wavefields propagating through the medium. (b) Graphic representation of the autocorrelation of the phase transmission in the situation depicted in (a), with P and S wavefields originating from the different kinks of the fibre (see equation (\ref{eq:plotting})).}
    \label{fig:fiber_green}
\end{figure}

Our objective is to compute the autocorrelation of the phase change of a light wave travelling through a fibre with a simplified geometry. This approach aims to both minimise computational complexity and enhance the interpretability of the autocorrelation results by reducing the number of kinks in the fibre. We use a triangle-shaped fibre, as shown in figure \ref{fig:fiber_green}. A small number of point sources simulates the generation of seismic noise.
We can define the directional sensitivity $\bm{a}(s)$ from equation (\ref{eq:implement}) for this fibre geometry as
\begin{multline}
    \bm{a}(s)= \frac{1}{2\Delta s} \biggl[ \left[ \frac{1}{\sqrt{2}}(\bm{e_1}-\bm{e_2}) + \bm{e_2}\right] \delta(s-s_1) +  \left[ \frac{-1}{\sqrt{2}}2\bm{e_1}\right]\delta(s-s_2) \\
    + \left[-\bm{e_2} - \frac{1}{\sqrt{2}}(-\bm{e_2}-\bm{e_1})\right]\delta(s-s_3) \biggr] 
\end{multline}
with $\bm{e}_1$ and $\bm{e}_2$ representing the unit vectors in $x$- and $y$-directions, respectively. Substituting $\bm{a}(s)$ into equation (\ref{eq:implement}) yields the following relationship, illustrated in figure \ref{fig:fiber_green}b,

\begin{equation}
\begin{aligned}
    c\left(\omega\right) \propto \textcolor{red}{G\left[\mathbf{x}\left(s_1\right),\mathbf{x}\left(s_1\right)\right] + G\left[\mathbf{x}\left(s_1\right),\mathbf{x}\left(s_2\right)\right] + G\left[\mathbf{x}\left(s_1\right),\mathbf{x}\left(s_3\right)\right]} \\
    + \textcolor{ForestGreen}{G\left[\mathbf{x}\left(s_2\right),\mathbf{x}\left(s_2\right)\right] + G\left[\mathbf{x}\left(s_2\right),\mathbf{x}\left(s_1\right)\right] + G\left[\mathbf{x}\left(s_2\right),\mathbf{x}\left(s_3\right)\right]} \\
    + \textcolor{gray}{G\left[\mathbf{x}\left(s_3\right),\mathbf{x}\left(s_3\right)\right] + G\left[\mathbf{x}\left(s_3\right),\mathbf{x}\left(s_1\right)\right] + G\left[\mathbf{x}\left(s_3\right),\mathbf{x}\left(s_2\right)\right]}
    \label{eq:plotting}
\end{aligned}
\end{equation}

where different colours of the Green's functions correspond to the wavefields propagating through the medium, as depicted by the matching colours in figure \ref{fig:fiber_green}b.

\begin{figure}
    \centering
    \includegraphics[width=1\linewidth]{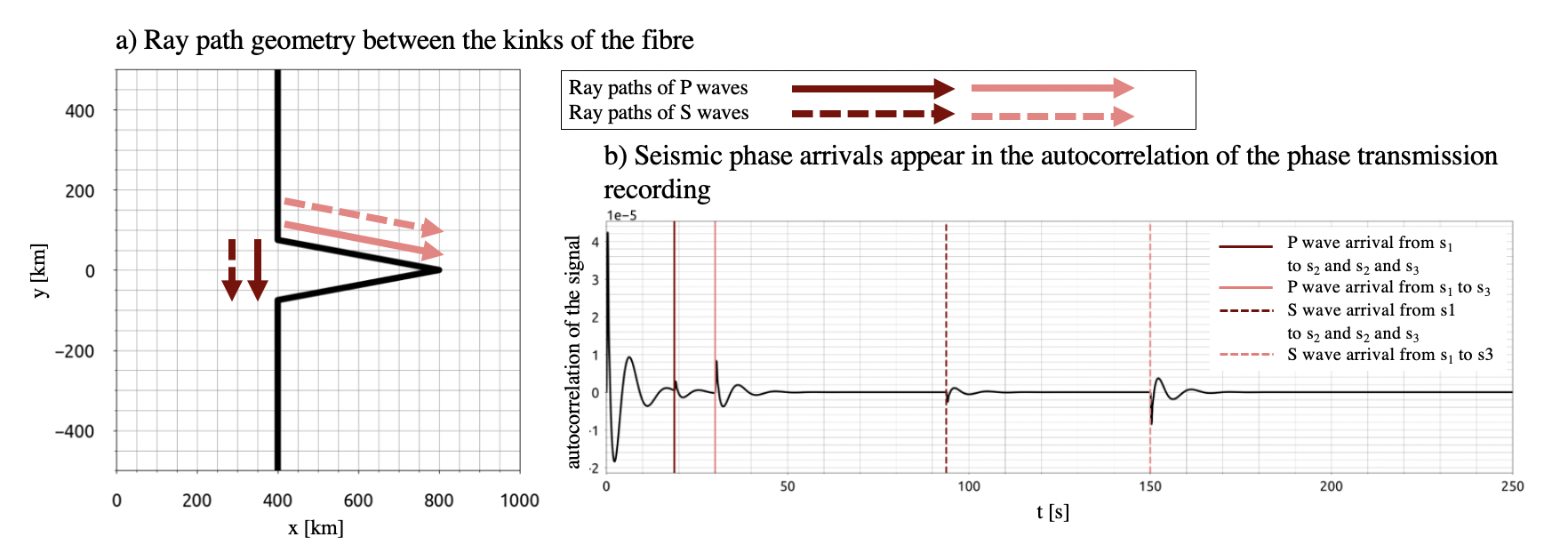}
    \caption[Synthetic seismogramm]{Computed autocorrelation for the conceptual example. (a) The fibre is represented in black. P and S wave ray paths are drawn as solid and dashed arrows, respectively. (b) Synthetic autocorrelation plotted against time with pulses in the autocorrelation corresponding to the arrival times of the P and S waves travelling from one fibre kink to another. The peak in the autocorrelation at time 0 s corresponds to the Green's functions of the P and S waves from sources that originate and are received at the same points.}
    \label{fig:fiber_seismo}
\end{figure}

To model equation (\ref{eq:plotting}) numerically, we consider the Green's function to be excited by point sources that radiate one P and one S wave, respectively, as illustrated in figure \ref{fig:fiber_seismo}. The medium in which the seismic waves propagate in our model is homogeneous, isotropic, unbounded and perfectly elastic. We then implement the autocorrelation equation (\ref{eq:plotting}), and plot it against time, as shown in figure \ref{fig:fiber_seismo}. The different spikes in the autocorrelation correspond to the arrival times of the P and S waves of the Green's functions at the points $s_1$, $s_2$ and $s_3$.
%
As the autocorrelation allows us to extract the travel-times corresponding to the arrival of seismic waves at a certain point in space, it in principle allows for seismic tomography, as one could infer the seismic velocities from the travel-times and the kink positions along the fibre. In the follow section we assess if useful autocorrelations may be extracted from actual phase transmission data.

\section{Case study : Fibre optic cable between Bern and Basel, Switzerland}

\begin{figure}[h]
    \centering
    \includegraphics[width=1\linewidth]{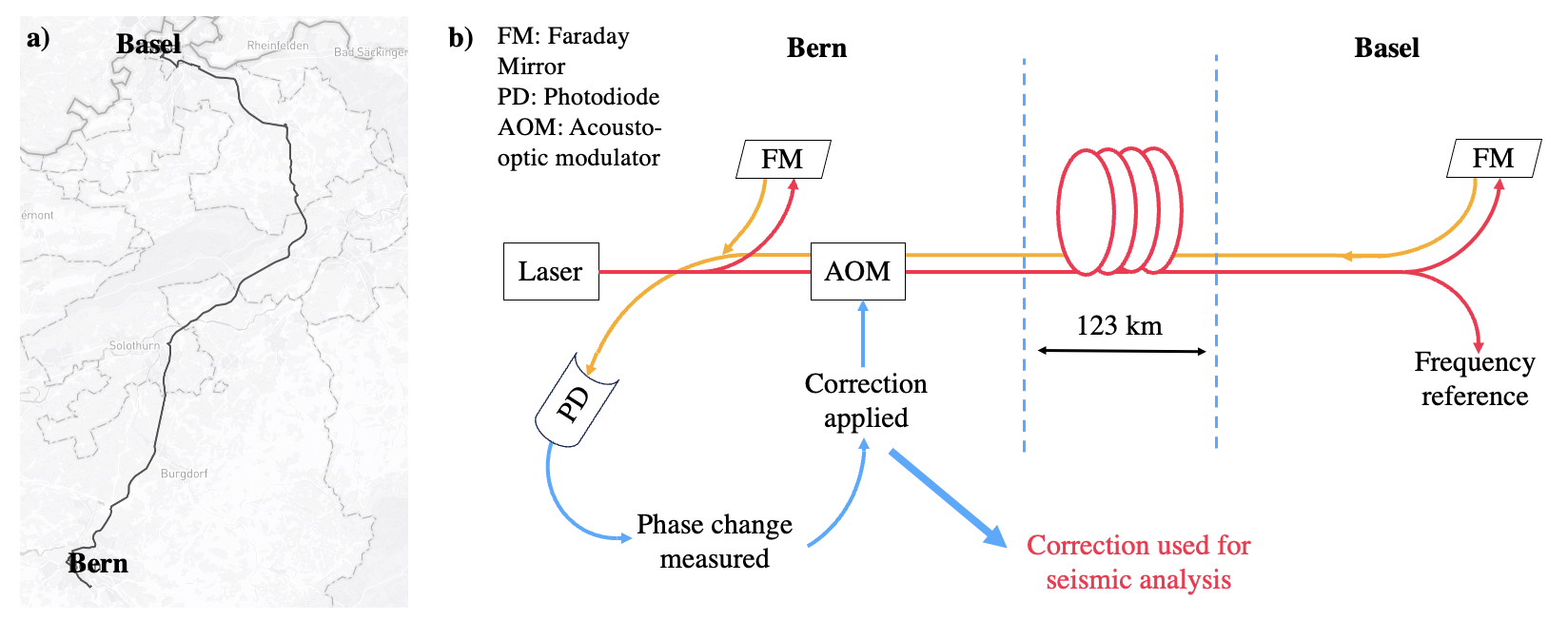}
    \caption[METAS set-up]{Setup of the phase transmission experiment. (a) Map of the fibre layout. (b) Simplified schematic of the phase measurement at METAS (Bern) and in Basel. An ultrastable laser signal at METAS is injected into the fibre network and transmitted to Basel over 123 km distance, with two inline bidirectional amplification stages (not shown) to compensate for fibre attenuation. In Basel, part of the light is coupled out for local frequency referencing. A portion of the signal is reflected back at the station in Bern by a Faraday mirror, while another portion is reflected back in Basel, forming a strongly imbalanced Michelson-type interferometer. The beat signal between the two interferometer arms thus contains the phase distortions picked up on the round-trip of the 123 km fibre, and is detected on a photodiode (PD). Subsequently a frequency correction signal is applied to the optical carrier via an acousto-optic modulator (AOM) at a 500 Hz sampling rate. This correction is what we will use for our seismic analysis.}
    \label{fig:METAS}
\end{figure}

The data analysed here were measured on a phase-stabilised fibre segment connecting the Swiss Federal Institute of Metrology (METAS) to the University of Basel. This is part of a larger fibre network operated by METAS, Switch, the University of Basel and ETH Zurich, which connects the SI-traceable frequency references at METAS to precision spectroscopy research groups in the connected institutions. To that end, an ultrastable optical frequency of 190.7 THz is generated at METAS and rendered SI-traceable by referencing to atomic clocks and the local realisation of Universal Time Coordinate, UTC(CH), which is traceable to international atomic time (TAI) via satellite comparisons \citep{Husmann:21}. Each of the fibre segments Bern-Basel, Basel-Zurich, and Zurich-Bern, are individually phase stabilised. Here, we only focus on the Bern-Basel segment, spanning 123 km of fibre distance \citep{noe2023long}. A schematic description of the set-up is shown in figure \ref{fig:METAS}.
As the light wave propagates through the fibre, mechanical strain induces phase shifts in the optical signal. These phase shifts must be corrected to ensure the accuracy of the optical reference frequency. Hence, the phase noise picked up by the fibre is coherently detected (see figure \ref{fig:METAS}b) and actively compensated by correcting the optical frequency via an acousto-optic modulator. Here we analyse this signal with respect to noise processes attributed to ground motion \citep{Husmann:21}.

The frequency analysis of phase noise cancellation (PNC) presented by \cite{Husmann:21} reveals that low noise levels are generally observed during nighttime, while daytime periods are characterised by elevated noise levels, particularly at higher frequencies. The majority of phase noise falls within the frequency range of $10$ to $100$ Hz, which is typically associated with anthropogenic sources \citep{nakata2019seismic}. However, certain components of the PNC exhibit frequencies between $0.1$ and $1$ Hz, which are typical for ocean-generated seismic ambient noise \citep{nakata2019seismic}.

\subsection{Synthetic analysis}

\begin{figure}[h]
    \centering
    \includegraphics[width=1\linewidth]{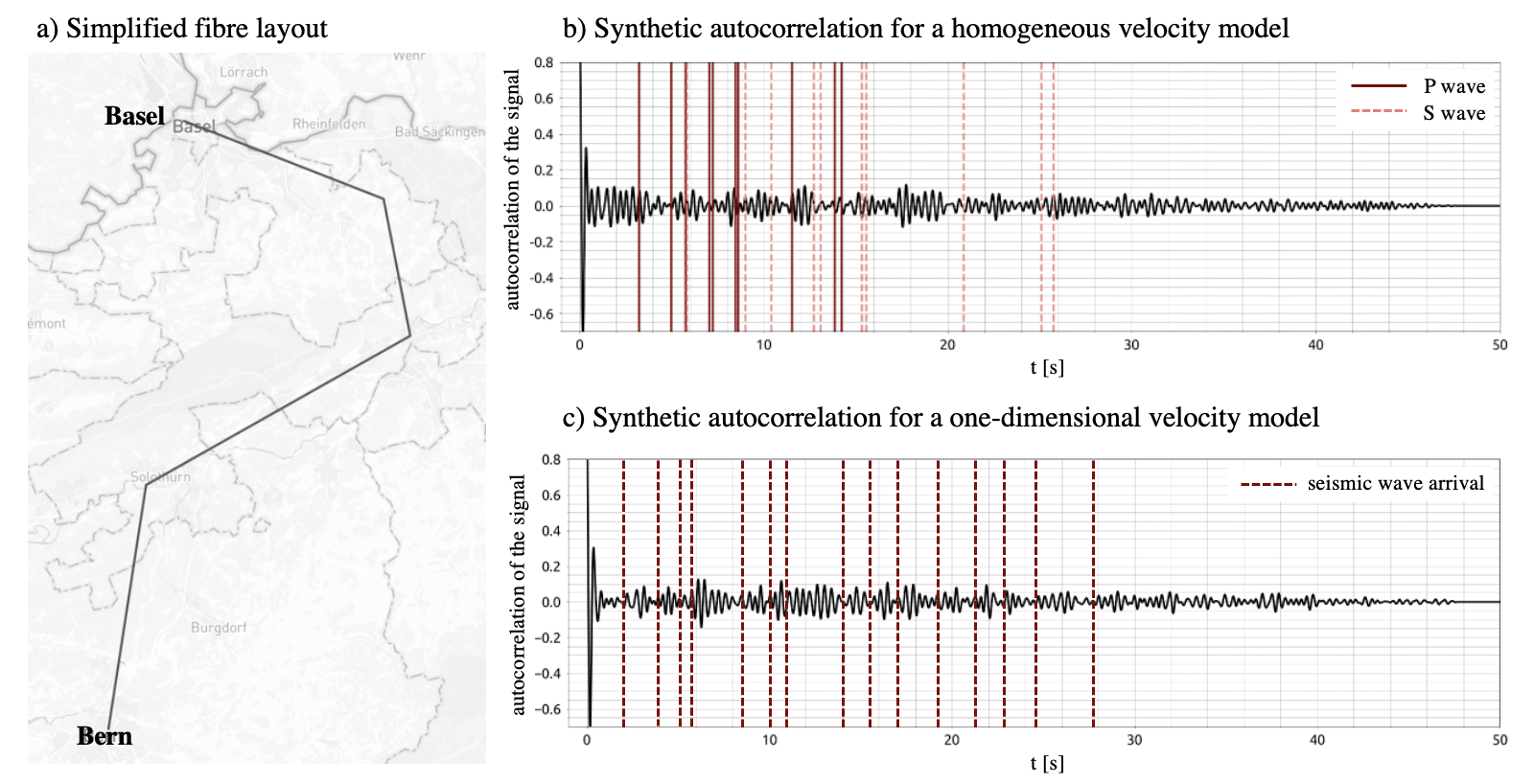}
    \caption[Simplified cable geometry and seismogramms]{Synthetic experiment. (a) Simplified layout of the cable. (b) Autocorrelation computed from the synthetic phase transmission with a homogeneous velocity model, with arrival times for the P and S waves marked by dashed lines. (c) Autocorrelation computed for a one-dimensional velocity model, with different seismic phase arrival times depicted. In (b) and (c), the peaks in the autocorrelations at time 0 s correspond to the Green's functions of the P and S waves of the sources that originate and are received at the same kinks.}
    \label{fig:simple_cable}
\end{figure}

Before turning to the real data, we first consider a simplified cable layout and numerically simulate the seismic wavefield, and its corresponding autocorrelation. Doing so will guide and inform us as to where we may expect to see distinct arrivals and be able to make travel-time measurements for seismic interferometry. We begin by simplifying our cable geometry into four linear segments, as shown in figure \ref{fig:simple_cable}a. We then calculate the integrated phase transmission that would be measured at the master station in Bern, following the methodology outlined by \cite{noe2023long}.

To model the seismic ambient noise wavefield, we simulate 10,000 point sources randomly in the domain near the surface, positioned approximately 20 km from the fibre. We use a Ricker wavelet as source time function in order to obtain a numerical approximation of a band-limited Green's function, with frequencies ranging from 0.1 to 0.66 Hz. In the first experiment, the domain is characterised by a homogeneous velocity model, while in the second experiment, a one-dimensional velocity model is employed, as illustrated in Figures \ref{fig:simple_cable}b and c. In both scenarios, the topography is neglected, assuming a flat domain.

We can observe distinct wave packets in both autocorrelations of the synthetic phase transmission in figures \ref{fig:simple_cable}b and c. The phases of the arrival times for the homogeneous velocity model are easier to interpret than the arrival times of the one-dimensional velocity model. We are indeed able for the first case to compute the expected arrival times as the P-wave velocity is set to 4800 m/s and the S-wave velocity is set to 2653 m/s (see table \ref{tab:interkink}). Nevertheless, we can still see that there are distinct peaks in the two autocorrelations in figures \ref{fig:simple_cable}b and c, as predicted by our simple analytical model in the previous section.

\begin{table}[]
\centering
\small
\begin{tabular}{|p{3.5cm}|p{4.5cm}|p{4.5cm}|}
\hline
\multicolumn{1}{|l|}{Interkink distances (in km)} & \multicolumn{1}{l|}{Expected P wave arrival times (in s)} & \multicolumn{1}{l|}{Expected S wave arrival times (in s)} \\ \hline
15.48                                             & 3.22                                                      & 5.83                                                      \\
23.88                                             & 4.98                                                      & 9.00                                                         \\
27.65                                             & 5.76                                                      & 10.42                                                     \\
33.78                                             & 7.04                                                      & 12.73                                                     \\
34.74                                             & 7.24                                                      & 13.10                                                     \\
40.66                                             & 8.47                                                      & 15.33                                                     \\
41.32                                             & 8.61                                                      & 15.57                                                     \\
55.36                                             & 11.53                                                     & 20.87                                                     \\
66.59                                             & 13.87                                                     & 25.10                                                     \\
68.28                                             & 14.23                                                     & 25.74   													\\ \hline                                                
\end{tabular}
\caption{Interkink distances and expected arrival times for known P and S waves, respectively 4800m/s and 2653m/s, and simplified cable geometry.}
\label{tab:interkink}
\end{table}


We now compute the autocorrelation for the actual fibre geometry as depicted in Figure \ref{fig:METAS}a. We keep the same configuration for the sources than for the previous calculations shown in figure \ref{fig:simple_cable}.

\begin{figure}[H]
    \centering
    \includegraphics[width=1\linewidth]{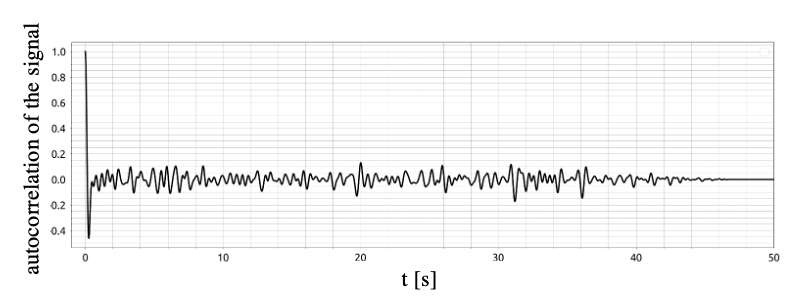}
    \caption[Complex cable geometry seismogramm]{Synthetic autocorrelation of the phase transmission calculated for the actual cable geometry}
    \label{fig:complex_cable}
\end{figure}

We can see that the autocorrelation is harder to interpret, as the kinks are too close to each other to actually be able to create separable pulses in the autocorrelation. This could be overcome by simulating sources with higher frequencies, but this would prove to be very computationally expensive on a domain of this size.

\subsection{Data analysis}

\begin{figure}
    \centering
    \includegraphics[width=1\linewidth]{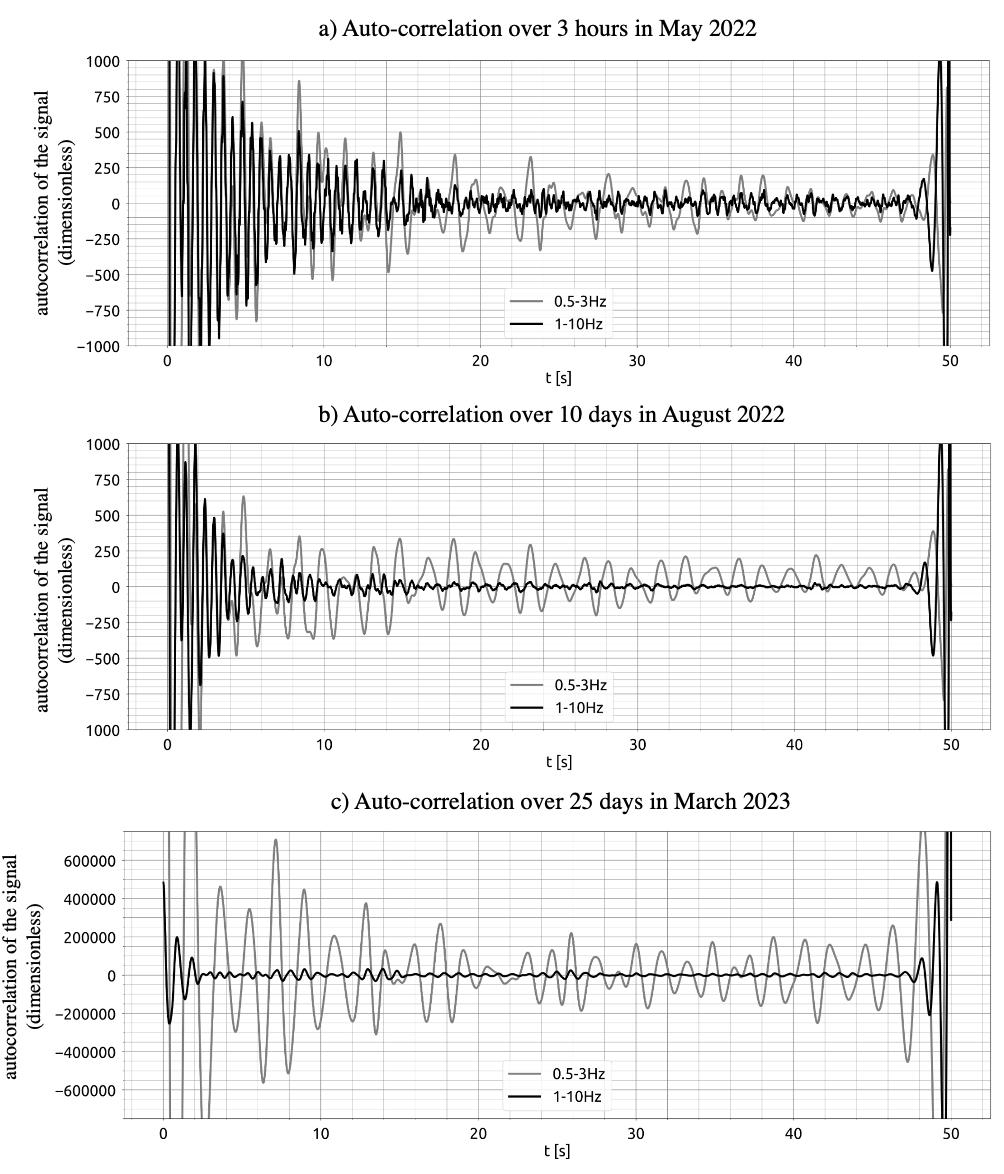}
    \caption[Data seismogramm]{Three autocorrelations of the PNC were computed over varying time periods. All datasets were processed using a Butterworth bandpass filter with frequency ranges of 0.5–3 Hz and 1–10 Hz, respectively. The data were segmented into one-hour intervals for autocorrelation analysis, and the results were subsequently averaged. A uniform sampling frequency of 50 Hz was applied across all datasets. The autocorrelations are dimensionless, as they have been normalised. (a) Autocorrelation over 3 hours. (b) Autocorrelation over 10 days. (c) Autocorrelation over 25 days. }
    \label{fig:data}
\end{figure}

We proceed to analyse the integrated PNC data provided by METAS, encompassing three distinct datasets, each varying in length and spanning different time intervals between 2022 and 2023. The autocorrelated PNC data have been processed using a Butterworth bandpass filter to isolate frequencies within the 0.5 to 3 Hz and 1 to 10 Hz ranges (see figure \ref{fig:data}). For normalisation, the autocorrelated PNC was divided by the signal variance. The primary focus is on the lower frequency range, as these frequencies are more likely to capture seismic signals, which undergo significant attenuation as they propagate through the Earth's interior. Beyond filtering, no further signal processing was applied.

As anticipated from the synthetic example shown in figure \ref{fig:complex_cable}, the autocorrelation is challenging to interpret. The expected peaks corresponding to the arrival times at the fibre kinks overlap due to their close proximity along the cable. Moreover, the comparison with synthetic data is highly sensitive to the one-dimensional local seismic velocity model used in the simulation. This model may not accurately represent the true seismic velocities, making it difficult to perform reliable seismic interferometry by comparing the observed and synthetic autocorrelations of phase-transmitted data. Additionally, the uncertainty in the exact cable location and geometry further complicates the analysis.

\section{Discussion and Conclusions}

Ambient noise interferometry through the recording of transmitted phase changes of an optical wave travelling through a fibre-optic cable is a new technology that still needs additional understanding beyond the scope of this article. Phase transmission fibre-optic sensing is a promising technology that could lead to interesting applications. While earthquake detection and location have already been demonstrated in practice \citep[e.g.,][]{noe2023long,mueller2024earthquake,donadello2024seismic}, also seismic monitoring and imaging using autocorrelations of noise recordings should be possible, as shown theoretically in this work. One interesting application would be to retrieve the phase transmission recording inside telecommunication cables placed on the ocean bottom, as proposed by \cite{marra2018ultrastable}. By performing interferometry of seismic noise recorded by the phase transmission recording, one could interpret in more detail Earth's internal structure in regions that lack detailed seismic tomography analysis due to insufficient seismometer distribution in oceanic regions in comparison to continental areas.

However, the application of this technique necessitates a fibre geometry which contains higher-curvature segments that are far apart from each other, and a precisely known fibre geometry, as the interpretation of the data is entirely dependent on it.  As discussed by \cite{marra2022optical}, one method to mitigate the uncertainty associated with cable geometry involves partitioning the optical fibre into discrete segments, each extending from one repeater to the next. This approach facilitates the interpretation of the autocorrelation function by transforming the problem: rather than the irregularities or kinks in the cable serving as sensors, the repeaters themselves function as measurement points.

Moreover, it is essential to quantify the fibre curvature, which would, for instance, characterise a kink as a discrete measurement point. In the same context, the displacement of a highly curved fibre-optic spool would only be detectable if the wavelength of the inducing wave is sufficiently large to prevent the simultaneous displacement of the entire spool.  

The coupling of the cable to the ground should be better constrained as well, as it determines the sensitivity of the cable to ground motion, an issue which has also been underlined by \cite{noe2023long}.

Currently, a dataset that fully meets these criteria has not yet been identified, making it challenging to provide a compelling case study that demonstrates the theoretical framework described in this article.

\subsection*{Acknowledgements}

This work represents a summary of the Bachelor Thesis written by Sixtine Dromigny at ETH Zurich. We thank the group of Prof. Stefan Willitsch and his group for continued support with the PNC installation in Basel. We acknowledge SWITCH for providing the fiber network infrastructure and its geo data, and Fabian Mauchle for technical support with the network, as well as many fruitful discussions with Daniel Bowden, Sixtine Dromigny, Pascal Edme, Sara Klaasen, Patrick Paitz, Krystyna Smolinski, Jacques Morel, Jerome Faist, Ernst Heiri, Fabian Mauchle, Ziv Meir, Frédéric Merkt, Giacomo Scalari and Stefan Willitsch. Funding was provided by the European Union’s Horizon 2020 research and innovation program under the Marie Sklodowska-Curie grant agreement No. 955515 (SPIN ITN), and by the Swiss National Science Foundation (SNSF) Sinergia grant CRSII5\_183579.

\bibliographystyle{apa-good.bst}
\bibliography{bib.bib}

\end{document}